\documentclass[12pt]{article}
\usepackage{jhep-mod}
\usepackage{bm}
\usepackage{amssymb,amsmath,amsthm}
\usepackage{mathrsfs}
\usepackage[utf8]{inputenc}
\usepackage{enumerate}
\usepackage{bigints}
\usepackage{appendix}
\usepackage{graphicx}
\usepackage{float}
\usepackage{tikz}
\usepackage{setspace}
\usepackage{cancel}
\definecolor{purple}{rgb}{1,0,1}
\definecolor{lime}{HTML}{A6CE39} 

\definecolor{lime}{HTML}{A6CE39}
\newcommand{\orcidicon}{%
	\begin{tikzpicture}
	\draw[lime, fill=lime] (0,0) 
		circle [radius=0.16] 
		node[white] {{\fontfamily{qag}\selectfont \tiny ID}};
	\draw[white, fill=white] (-0.0625,0.095) 
		circle [radius=0.007];
	\end{tikzpicture}
	\hspace{-5mm}
}
\newcommand\orcidAlex{{\href{https://orcid.org/0000-0002-1763-3563}{\orcidicon}}}
\newcommand\orcidMatt{{\href{https://orcid.org/0000-0003-1088-6485}{\orcidicon}}}
\begin{document}
\title{\vspace{-25pt}\huge{Decomposition of total stress-energy for the generalised Kiselev black hole}}
\author{
\Large Petarpa Boonserm$\,^{1,3}$, Tritos Ngampitipan$\,^{2,3}$, \\
Alex Simpson$\,^{4}$\orcidAlex\!\!, {\sf  and} Matt Visser$\,^{4}$\orcidMatt}
\affiliation{
$^{1}$  Department of Mathematics and Computer Science, Faculty of Science,\\
\null\qquad Chulalongkorn University,  Bangkok 10330, Thailand.  
}
\affiliation{
$^{2}$  Faculty of Science, Chandrakasem Rajabhat University,
 Bangkok 10900, Thailand.
}
\affiliation{$^{3}$ Thailand Center of Excellence in Physics, \\ \null\qquad 
Ministry of Higher Education, Science, Research and Innovation, \\ \null\qquad 
328 Si Ayutthaya Road, Bangkok 10400, Thailand.
}
\affiliation{$^{4}$ School of Mathematics and Statistics, Victoria University of Wellington, \\
\null\qquad PO Box 600, Wellington 6140, New Zealand.}
\emailAdd{petarpa.boonserm@gmail.com}
\emailAdd{tritos.ngampitipan@gmail.com}
\emailAdd{alex.simpson@sms.vuw.ac.nz}
\emailAdd{matt.visser@sms.vuw.ac.nz}

\abstract{
\\
We demonstrate that the anisotropic stress-energy supporting the Kiselev black hole can be mimicked by being split into a perfect fluid component plus either an electromagnetic component or a scalar field component, thereby quantifying the precise extent to which the Kiselev black hole fails to represent a perfect fluid spacetime. 
The perfect fluid component carries either an electric or a scalar charge, which then generates anisotropic electromagnetic or scalar fields. This in turn generates anisotropic contributions to the stress-energy. These in turn induce forces which partially (in addition to the fluid pressure gradient) support the matter content against gravity. 
This decomposition is carried out both for the original 1-component Kiselev black hole and for the generalized $N$-component Kiselev black holes. 
We also comment on the presence of energy condition violations (specifically for the null energy condition --- NEC) for certain sub-classes of Kiselev black holes.

\bigskip
\noindent
{\sc Date:} 18 October 2019; \LaTeX-ed \today


\bigskip
\noindent{\sc PhySH:} 
Gravitation; Classical black holes; \\
Fluids \& classical fields in curved spacetime.
}

\maketitle
\def\tr{{\mathrm{tr}}}
\def\diag{{\mathrm{diag}}}
\def\cof{{\mathrm{cof}}}
\def\pdet{{\mathrm{pdet}}}
\parindent0pt
\parskip7pt
\section{Introduction}\label{S:Intro}

The Kiselev black hole has proved to be an extremely popular and enduring toy model, with over 200 citations to date~\cite{Kiselev:2002}. Recently one of the current authors has pointed out that, despite many repeated claims to the contrary, the Kiselev black hole is not a perfect fluid spacetime, nor does it have anything to do with the notion of cosmological quintessence~\cite{Visser:2019}. 
(Relatively few articles in the follow-up literature are at all careful in this regard, a notable exception is reference~\cite{Cvetic:2016}.)
In the current article, we extend previous work by three of the current authors~\cite{Boonserm:2015}, wherein it was demonstrated that any arbitrary static anisotropic fluid sphere in general relativity can mimicked by a decomposition into (prefect fluid) + (electromagnetic field) + (scalar field) components, to the specific case of the Kiselev black hole. 
Note this is a mimicking procedure, designed to give insight into general features of the underlying physics --- we do not claim this is an identity. 

We find a trichotomy: Depending on the parameters describing the Kiselev black hole, either one has (prefect fluid) + (electromagnetic field) components, or one has pure cosmological constant, or one has (prefect fluid) + (scalar field) components.
We furthermore generalize this analysis to multi-component Kiselev black holes where the mass function $m(r)$ is described by a Puiseux series~\cite{Puiseux}. 

Furthermore, the Kiselev black holes, being surrounded by matter, can be viewed as examples of ``dirty'' black holes~\cite{dirty-1,dirty-2}. 
Depending on the specific parameters of the Kiselev geometry, we shall show that this matter can and often  does violate the null energy condition (NEC), with potentially serious implications in terms of instability and other unusual behaviour~\cite{Barcelo:2002,Friedman:1993,Flanagan:1996,Ford:1994,Hartman:2016,Roman:1986,Buniy:2005,Buniy:2006,Martin-Moruno:2013a,Martin-Moruno:2013b,Martin-Moruno:2015,Martin-Moruno:2017}. 

\section{Stress-energy for generic anisotropic fluids}\label{S:anisotropic}

In reference~\cite{Boonserm:2015} it was established that one may ``mimic'' the total stress-energy of a general anisotropic fluid sphere by a combination of
\[
\hbox{ (perfect fluid) + (electromagnetic field) + (massless minimally coupled scalar field). }
\]
That is:
\begin{equation}\label{E:general-decomp}
    T^{\hat{a}\hat{b}}_{total} = T^{\hat{a}\hat{b}}_{f} + T^{\hat{a}\hat{b}}_{em} + T^{\hat{a}\hat{b}}_{s}.
\end{equation}
Here the general forms for the various stress-energy tensors are 
\begin{equation}
    T^{\hat{a}\hat{b}}_{total} = 
    \begin{bmatrix}
        \rho & 0 & 0 & 0 \\
        0 & p_{r} & 0 & 0 \\
        0 & 0 & p_{t} & 0 \\
        0 & 0 & 0 & p_{t}
    \end{bmatrix};
    \qquad
    T^{\hat{a}\hat{b}}_{f} = 
    \begin{bmatrix}
        \rho_{f} & 0 & 0 & 0 \\
        0 & p_{f} & 0 & 0 \\
        0 & 0 & p_{f} & 0 \\
        0 & 0 & 0 & p_{f}
    \end{bmatrix}; \nonumber \\
   \end{equation}
   and
   \begin{equation}
    T^{\hat{a}\hat{b}}_{em} = \frac{1}{2}E^{2}
    \begin{bmatrix}
        +1 & 0 & 0 & 0 \\
        0 & -1 & 0 & 0 \\
        0 & 0 & +1 & 0 \\
        0 & 0 & 0 & +1
    \end{bmatrix};
    \qquad
    T^{\hat{a}\hat{b}}_{s} = \frac{1}{2}\left(\nabla\phi\right)^{2}
    \begin{bmatrix}
        +1 & 0 & 0 & 0 \\
        0 & +1 & 0 & 0 \\
        0 & 0 & -1 & 0 \\
        0 & 0 & 0 & -1
    \end{bmatrix}. \qquad 
\end{equation}
\enlargethispage{20pt}
The perfect fluid parameters are given by
\begin{equation}
    p_{f} = \frac{1}{2}\left(p_{r}+p_{t}\right); \qquad\qquad
    \rho_{f} = \rho -\frac{1}{2}\vert p_{r}-p_{t}\vert. \\
\end{equation}  
 The electromagnetic/scalar field parameters are given by
 \begin{equation}   
    E^{2} = \ \mbox{max}\lbrace p_{t}-p_{r}, 0\rbrace;\qquad\qquad
    \left(\nabla\phi\right)^{2} = \ \mbox{max}\lbrace p_{r}-p_{t}, 0\rbrace.
\end{equation}

Already at this level we encounter a trichotomy: At any particular value of the radial coordinate $r$, either one has (prefect fluid) + (electromagnetic field) components, or one has pure perfect fluid, or one has (prefect fluid) + (scalar field) components.
Let us now perform this decomposition in more detail for the specific case of the Kiselev black hole.

\section{Stress-energy of the 1-component Kiselev black hole}\label{S:1-component}

The (1-component) Kiselev black hole spacetime is defined by the line element~\cite{Kiselev:2002}:
\begin{equation}
    ds^2 = -\left(1-\frac{2m}{r}-\frac{K}{r^{1+3w}}\right) \ dt^2 + \frac{dr^2}{1-\frac{2m}{r}-\frac{K}{r^{1+3w}}} + r^2 \ d\Omega_{2}^{2}.
\end{equation}

The Einstein tensor components with respect to an orthonormal basis are:
\begin{equation}
    G_{\hat{t}\hat{t}} = -G_{\hat{r}\hat{r}} = -\frac{3Kw}{r^{3\left(1+w\right)}}; \qquad 
    G_{\hat{\theta}\hat{\theta}} = G_{\hat{\phi}\hat{\phi}} = -\frac{3Kw\left(1+3w\right)}{2r^{3\left(1+w\right)}}.
\end{equation}

Consequently the stress-energy tensor components are
\begin{eqnarray}
    \rho = -p_{r} = -\frac{3Kw}{8\pi r^{3\left(1+w\right)}}; \qquad 
    p_{t} = -\frac{3Kw\left(1+3w\right)}{16\pi r^{3\left(1+w\right)}}.
\end{eqnarray}

Examining
\begin{equation}
    p_{r}-p_{t} = -\left(p_{t}-p_{r}\right) = \frac{9Kw\left(1+w\right)}{16\pi r^{3\left(1+w\right)}},
\end{equation}
we see that either $\left(\nabla\phi\right)^{2}=0$ or $E^{2}=0$, in a position-independent manner, subject only to the \emph{sign} of the constant $Kw(1+w)$. 
\enlargethispage{30pt}
We also note
\begin{equation}
\rho+p_r=0; \qquad \rho+p_t = -\frac{9Kw\left(1+w\right)}{16\pi r^{3\left(1+w\right)}},
\end{equation}
so the null energy condition (NEC) is either satisfied or violated depending on the \emph{sign} of the constant $Kw(1+w)$. 
(For background on the classical and semi-classical energy conditions see references~\cite{Barcelo:2002,Friedman:1993,Flanagan:1996,Ford:1994,Hartman:2016,Roman:1986,Buniy:2005,Buniy:2006,Martin-Moruno:2013a,Martin-Moruno:2013b,Martin-Moruno:2015,Martin-Moruno:2017}.)
We now develop a fully-explicit case-by-case argument, based on the \emph{sign} of the constant $Kw(1+w)$, to determine the specific form of the  linear decomposition presented in equation~(\ref{E:general-decomp}).

\subsection{Case (i): $Kw(1+w)<0$. NEC satisfied.}
In this situation we have $p_{r}-p_{t}<0$. Therefore $\left(\nabla\phi\right)^{2}=0$. We obtain:
\begin{eqnarray}
    p_{f} &=& \frac{1}{2}\left(p_{r}+p_{t}\right) = \frac{3Kw(1-3w)}{32\pi r^{3(1+w)}}; \\
    && \nonumber \\
    \rho_{f} &=& \rho - \frac{1}{2}\vert p_{r}-p_{t}\vert = -\frac{3Kw(1-3w)}{32\pi r^{3(1+w)}};\\
    && \nonumber \\
    w_f &=& {p_f\over\rho_f} = -1.
\end{eqnarray}
For the electromagnetic contribution $T^{\hat{a}\hat{b}}_{em}$ to the stress-energy we evaluate $E^2$:
\begin{equation}
    E^{2} = \ \mbox{max}\lbrace p_{t}-p_{r}, 0\rbrace = -\frac{9Kw(1+w)}{16\pi r^{3(1+w)}}.
\end{equation}
Therefore
\begin{eqnarray}
    T^{\hat{a}\hat{b}}_{f} = \frac{3Kw(1-3w)}{32\pi r^{3(1+w)}} \ \mbox{diag}\left(-1, +1, +1, +1\right); \\
    && \nonumber \\
    T^{\hat{a}\hat{b}}_{em} = -\frac{9Kw(1+w)}{32\pi r^{3(1+w)}} \ \mbox{diag}\left(+1, -1, +1, +1\right),
\end{eqnarray}
and we have the linear decomposition:
\begin{equation}
    T^{\hat{a}\hat{b}}_{total}  = T^{\hat{a}\hat{b}}_{f} + T^{\hat{a}\hat{b}}_{em} 
    = -\frac{3Kw}{8\pi r^{3(1+w)}} \ \mbox{diag}\left(+1, -1, \frac{1+3w}{2}, \frac{1+3w}{2}\right).
\end{equation}
Examining our expression for electric field strength $E(r)$ we have:
\begin{equation}
    E(r) = \pm\frac{3\sqrt{\vert Kw(1+w)\vert}}{4\sqrt{\pi}\; r^{3(1+w)/2}};
    \qquad 
    \frac{dE}{dr} = -\frac{3E(1+w)}{2r}.
\end{equation}

Invoking  Gauss' law, the charge inside a sphere of radius $r$ is
\begin{equation}
    Q(r) = E(r) \ 4\pi r^{2} =  \pm\frac{3\sqrt{\pi\vert Kw(1+w)\vert}}{{r^{(3w-1)/2}}}.
\end{equation}
We now compute $\sigma_{em}(r)$, the electric charge density
\begin{equation}
\sigma_{em} = \frac{dQ(r)}{dV} 
= \frac{dQ}{4\pi r^{2}\sqrt{g_{rr}} \ dr} 
= \frac{\sqrt{g^{rr}}}{4\pi r^{2}} \ \frac{d}{dr}\left[4E\pi r^{2}\right] 
= \sqrt{g^{rr}}\left[\frac{dE}{dr}+\frac{2E}{r}\right].
\end{equation}
Consequently
\begin{equation}
\sigma_{em}(r) =\sqrt{g^{rr}}\left[\frac{E(1-3w)}{2r}\right]
= \sqrt{1-{2m\over r} - {K\over r^{1+3w}}} \; \left[\frac{E(1-3w)}{2r}\right].
\end{equation}
To be fully explicit
\begin{equation}
\sigma_{em}(r) =\pm (1-3w) \sqrt{1-{2m\over r} - {K\over r^{1+3w}}} \times
\frac{3\sqrt{\vert Kw(1+w)\vert/\pi}}{8\; r^{1+3(1+w)/2}}.
\end{equation}
Note that $w=1/3$ is special in that the (distributed) charge density is zero --- indeed setting $w=1/3$ and $K\to-Q^2$ reproduces  Reissner--Nordstr\"om spacetime, with all electric charge concentrated at the origin.

\enlargethispage{20pt}
\subsection{Case (ii): $Kw(1+w)=0$. NEC marginal.}
This is the straightforward case. Either $K=0$ (standard Schwarzschild spacetime), or $w=0$ (also Schwarzschild spacetime but with shifted mass $m\rightarrow m+\frac{K}{2}$), or $w=-1$ (Kottler spacetime~\cite{Kottler:1918}, also called  Schwarzschild-(anti)-de Sitter spacetime). All three possibilities have $p_{r}-p_{t}=0$, \emph{i.e.} $\left(\nabla\phi\right)^{2} = E^{2} = 0$, so our decomposition of the total stress-energy must solely consist of the perfect fluid component of form $T^{\hat{a}\hat{b}}_{f}$. (The fact that these three spacetimes model a perfect fluid is of course extremely standard.)

By inspection, if either $K$ or $w=0$ then we have the trivial linear decomposition:
\begin{equation}
    T^{\hat{a}\hat{b}}_{total} = T^{\hat{a}\hat{b}}_{f} = \mbox{diag}(0,0,0,0),
\end{equation}
reflective of the fact that Schwarzschild is a vacuum solution to the Einstein equations.

If $w = -1$ then we have:
\begin{equation}
    \rho_{f} = \rho = \frac{3K}{8\pi} \ ; \qquad p_{f} = \frac{1}{2}(p_{r}+p_{t}) = -\frac{3K}{8\pi} \ ; \qquad p_{r}=p_{t}=-\frac{3K}{8\pi} \ .
\end{equation}
So our linear decomposition is (also rather trivially):
\begin{equation}
    T^{\hat{a}\hat{b}}_{total} = T^{\hat{a}\hat{b}}_{f} = \frac{3K}{8\pi}\mbox{diag}(+1, -1, -1, -1) \ .
\end{equation}

\subsection{Case (iii): $Kw(1+w)>0$. NEC violated.}
In this situation we have $p_{t}-p_{r}<0$. Therefore $E^{2}=0$, and we obtain:
\begin{eqnarray}
    p_{f} &=& \frac{1}{2}(p_{r}+p_{t}) = \frac{3Kw\left(1-3w\right)}{32\pi r^{3(1+w)}}; \\
    && \nonumber \\
    \rho_{f} &=& \rho - \frac{1}{2}\vert p_{r}-p_{t}\vert = -\frac{3Kw\left(7+3w\right)}{32\pi r^{3(1+w)}};\\
    && \nonumber \\
    w_{f} &=& \frac{p_{f}}{\rho_{f}} = -\;\frac{1-3w}{7+3w}.
\end{eqnarray}
Note that in this situation
\begin{equation}
\rho_{f} +p_f =  -\frac{9Kw\left(1+w\right)}{16\pi r^{3(1+w)}} < 0,
\end{equation}
implying NEC violation for the perfect fluid component. While this NEC violation is certainly disturbing, let us nevertheless carry out the case (iii) analysis as far as we can.

For the scalar field contribution $T^{\hat{a}\hat{b}}_{s}$ we evaluate:
\begin{equation}
    \left(\nabla\phi\right)^{2} = \mbox{max}\lbrace p_{r}-p_{t}, 0\rbrace = \frac{9Kw\left(1+w\right)}{16\pi r^{3(1+w)}}.
\end{equation}
Therefore:
\begin{eqnarray}
    T^{\hat{a}\hat{b}}_{f} &=& \frac{3Kw}{32\pi r^{3(1+w)}} \ \mbox{diag}\left(-3w-7, 1-3w, 1-3w, 1-3w\right); \\
    && \nonumber \\
    T^{\hat{a}\hat{b}}_{s} &=& \frac{9Kw(1+w)}{32\pi r^{3(1+w)}} \ \mbox{diag}\left(+1, +1, -1, -1\right);
\end{eqnarray}
and 
\begin{equation}
     T^{\hat{a}\hat{b}}_{total}= T^{\hat{a}\hat{b}}_{f} + T^{\hat{a}\hat{b}}_{s} 
     = -\frac{3Kw}{8\pi r^{3(1+w)}} \ \mbox{diag}\left(+1, -1, \frac{1+3w}{2}, \frac{1+3w}{2}\right).
\end{equation}

Solving explicitly for our scalar field, $\phi(r)$ we have
\begin{equation}
\left(\nabla\phi\right)^{2} = g^{ab}\;\partial_{a}\phi \; \partial_{b}\phi 
= g^{rr}\left(\partial_r\phi\right)^{2} = \frac{9Kw(1+w)}{16\pi r^{3(1+w)}}.
\end{equation}
Consequently
\begin{equation}
\left(\partial_r\phi\right)^{2} = g_{rr}\left(\frac{9Kw(1+w)}{16\pi r^{3(1+w)}}\right)
= {1\over{1-{2m\over r} - {K\over r^{1+3w}}}} \left(\frac{9Kw(1+w)}{16\pi r^{3(1+w)}}\right).
\end{equation}
Thence the scalar field $\phi(r)$ is given by
\begin{equation}
\phi(r) = \pm\frac{3}{4} \sqrt{Kw(1+w)/\pi}\bigintsss{dr\over\sqrt{1-{2m\over r} - {K\over r^{1+3w}}}\;\;  r^{3(1+w)/2}  }.
\end{equation}
Evaluating this integral directly is not analytically feasible.  

Instead we shall examine the scalar charge density:
\begin{eqnarray}
    \sigma_{s}(r) &=& \Delta\phi = \nabla^{2}\phi 
    = \frac{1}{\sqrt{-\det(g)}} \ \partial_{a}\left(\sqrt{-\det(g)} \; g^{ab} \; \partial_{b}\phi\right).
\end{eqnarray}
That is 
\begin{equation}
\sigma_{s}(r) = \frac{1}{r^{2}} \ \partial_{r}\left(r^{2} \; g^{rr} \; \partial_{r}\phi\right).
\end{equation}
From this we find an explicit but clumsy formula for the scalar charge density:
\begin{eqnarray}
    \sigma_{s}(r)
    &=& 
    \pm\frac{3}{4} {\sqrt{Kw(1+w)/\pi}\over r^{2}} \; 
    \partial_{r}\left({\textstyle\sqrt{1-{2m\over r} - {K\over r^{1+3w}} }}\;\; r^{(1-3w)/2} \right).
\end{eqnarray}

The total scalar charge inside a sphere of radius $r$ is then
\begin{equation}
S(r) =\int\sigma_{s}(r)\; dV = \int\sigma_{s}(r)\sqrt{g_{rr}}\;4\pi r^{2} \; dr.
\end{equation}
Explicitly 
\begin{equation}
S(r) = \pm{3} {\sqrt{Kw(1+w)\pi}} \; 
    \bigintsss_0^r   {1\over\sqrt{1-{2m\over r} - {K\over r^{1+3w}}}} \;
    \partial_{r}\!\left({\textstyle{\sqrt{1-{2m\over r} - {K\over r^{1+3w}} }}}\;\; r^{(1-3w)/2} \right) dr.
\end{equation}
While evaluating this integral directly is not analytically feasible, it is at least a fully explicit formula for the scalar charge $S(r)$. 
In short, case (iii) shares the two features of being physically dubious, (violating the NEC), and being technically clumsy to work with. 

\subsection{Summary (1-component model)}
It is worth explicitly pointing out here that an immediate corollary of the above case-by-case analysis is that the 1-component Kiselev spacetime is only ever a perfect fluid if we are dealing with case (ii); that is $Kw(1+w)=0$. Thus the Kiselev black hole is a perfect fluid spacetime only when it reduces to either Schwarzschild or Kottler spacetime~\cite{Kottler:1918}. The fact that the Kiselev solution does not generally model a perfect fluid is stressed in~\cite{Visser:2019}, and is an important point to reiterate in view of the quite common historical tendency to mis-identify perfect fluid models~\cite{Delgaty:1998}. (For more discussion of the constraints implied by imposing the perfect fluid condition see also references~\cite{Rahman:2001,Martin:2003a,Martin:2003b,Boonserm:2005,Boonserm:2006a,Boonserm:2006b,Boonserm:2007}.) 

Insofar as one is willing to accept the classical energy conditions as a pragmatic guideline~\cite{Barcelo:2002}, 
case (i) is the most physically interesting situation --- it corresponds to an electrically charged fluid supported by both pressure gradients and its own internally generated electric field. Case (ii) is physically more prosaic, representing either Schwarzschild or Kottler spacetime. Case (iii) is physically dubious and technically clumsy, exhibiting null energy condition violations. 

\section{Multi-component Kiselev decomposition}\label{S:N-component}

\enlargethispage{20pt}
Now consider the $N$-component generalization of Kiselev spacetime as presented in~\cite{Kiselev:2002}
\begin{equation}
ds^2 = - \left(1 - {\sum_{i=0}^N K_i \,r^{-3w_i}\over r} \right) dt^2 
+ {dr^2\over 1 - {\sum_{i=0}^N K_i \,r^{-3w_i}\over r}}
+ r^2 \,d\Omega_2^2.
\end{equation}
Any Schwarzschild mass term, if present, has now been absorbed into $K_0=2m$; while setting the corresponding exponent $w_0$ to zero.
That is, effectively one is defining a position-dependent mass function $m(r)$ by setting~\cite{Visser:2019}
\begin{equation}
\label{E:mass}
2\, m(r) = \sum_{i=0}^N K_i \,r^{-3w_i}.
\end{equation}
Since the exponents $w_i$ can be arbitrary, this is a Puiseux series expansion for the mass function, a generalization of the notion of Taylor series, Laurent series, power series, and Frobenius series~\cite{Puiseux}. Puiseux expansions, while somewhat uncommon, do have a number of other uses in astrophysical situations~\cite{Cattoen:2005,Cattoen:2006,Cattoen:2007,Visser:2002}.

The spacetime metric is then written in the form
\begin{equation}
\label{E:normal}
ds^2 = - \left(1 - {2m(r)\over r} \right) dt^2 
+ {dr^2\over 1 - {2m(r)\over r}}
+ r^2 \,d\Omega_2^2.
\end{equation}
Spacetime metrics of this form have very special properties~\cite{Jacobson:2007}.
For instance the radial coordinate $r$ acts as an affine parameter for radial null curves, the radial null-null components of the Einstein and Ricci tensors vanish~\cite{Jacobson:2007}, and the Einstein and Ricci tensors possess two Lorentz-invariant eigenvalues each of multiplicity two, so that the characteristic polynomial factorizes with two repeated roots, implying a specialized Rainich form for the stress-energy~\cite{Rainich}.  Spacetimes of this general form have also been extensively investigated by Dymnikova~\cite{Dymnikova:1999,Dymnikova:2000,Dymnikova:2003,Dymnikova:2004,Dymnikova:2010}.

It is an utterly standard calculation to show
\begin{equation}
\rho = - p_r = {m'(r)\over4\pi r^2},
\qquad
\hbox{and}
\qquad
p_t = -{m''(r)\over8\pi r}.
\end{equation}
Now note 
\begin{equation}
p_r-p_t = -(p_t-p_r) = { - 2m'(r)+ r m''(r)\over8\pi r^2} =  {r\over2} \left( m'\over4\pi r^2\right)' =  \; {r \over2}\; \rho'(r).
\end{equation}
Therefore in this multi-component Kiselev geometry it is the \emph{sign} of the density gradient $\rho'(r)$ that determines whether one is dealing with electromagnetic or scalar fields.  In contrast to the 1-component Kiselev geometry this may now change sign at various values of the radial coordinate leading to an onion-like layered object.  Note that
\begin{equation}
{p_r+p_t\over2} = -\rho - {p_r-p_t\over2} = - \rho - \; {r \over4}\; \rho'.
\end{equation}
Furthermore
\begin{equation}
\rho+p_r=0; \qquad \rho+p_t = - \; {r \over2}\; \rho'(r),
\end{equation}
so the null energy condition (NEC) is either satisfied or violated depending on the \emph{sign} of the density gradient $\rho'(r)$. 
We again perform a case-by-case analysis now conditioned on the sign of the density gradient.

\subsection{Case (i): $\rho'(r)<0$. NEC satisfied.}

In this situation $(\nabla\phi)^2=0$ while
\begin{equation}
\rho_f = \rho- {r \over4}\; |\rho'|; \qquad p_f = - \rho + \; {r \over4}\; |\rho'|; \qquad
w_f={p_f\over\rho_f}=-1; 
\qquad E^2 = {r \over2}\; |\rho'|. 
\end{equation}
Note that, as for the 1-component model, $w_f=-1$ in this situation. Furthermore
\begin{equation}
\rho_f + p_f =0,
\end{equation}
so the perfect fluid component marginally satisfies the NEC.

The electric charge inside a sphere of radius $r$ is now
\begin{equation}
    Q(r) = E(r) \; 4\pi r^{2} =  \pm 2\sqrt{2}\pi \; r^{5/2} \; \sqrt{ |\rho'|}.
\end{equation}
For the electric charge density
\begin{equation}
\sigma_{em}(r) = \frac{dQ(r)}{dV} = \sqrt{1-2m(r)/r}\left[\frac{dE}{dr}+\frac{2E}{r}\right].
\end{equation}
While we can make these formulae fully explicit in terms of $m(r)$, the discussion above is enough to clarify the basic physics issues. 

\subsection{Case (ii): $\rho'(r)=0$. NEC marginal.}

The situation $\rho'(r)=0$ could either arise ``instantaneously'' at the transition layer between $\rho'(r)<0$ (electromagnetic mimic) and $\rho'(r)>0$ (scalar field mimic), or it could hold over some finite interval of $r$. If $\rho'(r)=0$ holds over some finite interval, then $\rho(r) = \rho_*$ is a constant over that interval, and so $m(r) = m_* + {4\pi \over 3}  \rho_* r^3$ on that interval, 
so the spacetime is Kottler over that interval --- this corresponds to this region being described by a cosmological constant $\rho_\Lambda = \rho_*$. This is completely compatible with what we saw for the 1-component model.

\subsection{Case (iii): $\rho'(r)>0$. NEC violated.}
In this situation $E^2=0$ while we now have
\begin{equation}
\rho_f = \rho- {r \over4}\; \rho'; \qquad p_f = - \rho - \; {r \over4}\; \rho'; \qquad
w_f={p_f\over\rho_f}=-{\rho+{r\over4}\rho'\over\rho-{r\over4}\rho'}\neq -1; 
\end{equation}
Note that, as for the 1-component model, $w_f\neq-1$ in this situation. Furthermore
\begin{equation}
\rho_f + p_f = - {r \over2}\; \rho' < 0,
\end{equation}
implying NEC violation for the perfect fluid component.
(It should also be said that any ``dirty'' black hole for which the total energy density increases as one moves away from the centre is somewhat ``odd''.) 
Despite the NEC violations, let us push this analysis a little further to see how far we can get. 

For the scalar field we have
\begin{equation}
\qquad (\nabla\phi)^2 = {r \over2}\; \rho'. 
\end{equation}
Therefore, we see that
\begin{equation}
\partial_r \phi = \sqrt{g_{rr}} \;\; \nabla_{\hat r} \phi =  \sqrt{r \rho'/2\over 1-2m(r)/r},
\end{equation}
and so
\begin{equation}
\phi(r) = \bigintsss \sqrt{r \rho'/2\over 1-2m(r)/r} \; dr.
\end{equation}
The scalar charge density is
\begin{eqnarray}
    \sigma_{s}    &=& 
    \frac{1}{r^{2}} \ \partial_{r}\left(r^{2} \; g^{rr} \; \partial_{r}\phi\right)
    =
     \frac{1}{r^{2}} \ \partial_{r}\left(r^{2} \; \sqrt{g^{rr}} \; \nabla_{\hat r}\phi\right)
     \nonumber\\
     &=& 
     \frac{1}{r^{2}} \ \partial_{r}\left(r^{2} \; \sqrt{1-2m(r)/r}\; \sqrt{r \rho'/2} \right).
\end{eqnarray}
Finally the total scalar charge inside a sphere of radius $r$ is
\begin{equation}
S(r) = \int\sigma_{s}(r) \; dV =  \int{\sigma_{s}(r)\over\sqrt{1-2m(r)/r}}\;4\pi r^{2} \; dr. 
\end{equation}
While we can make these formulae fully explicit in terms of $m(r)$, the discussion above is enough to clarify the basic physics issues. 

\clearpage
\subsection{Summary (multi-component models)}

From the above, the only situation in which the generalized Kiselev spacetime represents a perfect fluid is in case (ii) when $\rho'=0$. That is, when the generalized Kiselev black hole reduces to Kottler (or Schwarzschild) spacetime.
The major difference between simple 1-component Kiselev spacetimes and these generalized multi-component Kiselev spacetimes is that the presence or absence of electromagnetic or scalar fields in the mimicking model can now depend on the radial coordinate $r$ in multi-component models,  whereas in 1-component models, there is only either electromagnetic  or scalar field everywhere or there is nothing (neither electromagnetic nor scalar field).

We again see violations of the NEC in this now generalized case (iii), and insofar as one wishes to be guided by the classical energy conditions, case (iii) should be deprecated. At the very least, if one wishes to work with case (iii) models, one should be aware of the potential risks and drawbacks.

\section{Discussion}\label{S:discussion}

While the Kiselev black hole~\cite{Kiselev:2002} is an extremely popular toy model, there are a number of key scientific  issues regarding  which the published literature is seriously deficient:
\begin{itemize}
\item Despite many claims to the contrary, the Kiselev spacetime does not represent a perfect fluid, (except for the very special cases where it reduces to Schwarzschild/\-Kottler/de~Sitter spacetime)~\cite{Visser:2019}. 
\item Despite many claims to the contrary, the word ``quintessence'' as applied to the Kiselev spacetimes has nothing to do with the word ``quintessence'' as it is used in the cosmology community~\cite{Visser:2019}. 
\end{itemize}
In earlier work three of the current authors showed that it is possible to mimic the matter content of any static spherically symmetric spacetime by a combination of
\[
\hbox{ (perfect fluid) + (electromagnetic field) + (scalar field). }
\]
In the current work we apply this decomposition to the specific case of the Kiselev spacetimes, both the original 1-component model and the generalized multi-component models. We find that there is a tight correlation between satisfying the null energy condition (NEC) and the type of decomposition that arises. 

We find a trichotomy:
\begin{itemize}
\item In regions where the Kiselev spacetime strongly satisfies the NEC its matter content can be mimicked by (electrically charged perfect fluid) + (electromagnetic field).
\item In regions where the Kiselev spacetime marginally satisfies the NEC its matter content is forced to be cosmological constant.
\item In regions where the Kiselev spacetime violates the NEC its matter content can awkwardly be mimicked by (scalar charged perfect fluid) + (scalar field).
\end{itemize}
Overall we would argue that while the Kiselev spacetime and its generalizations are certainly physically and mathematically interesting, 
some significant caution should be exercised when interpreting much of the current literature on Kiselev spactimes (and the Rastall\-ization thereof~\cite{Rastallization}).  

\acknowledgments{
This project was funded by the Ratchadapisek Sompoch Endowment Fund, Chulalongkorn University (Sci-Super 2014-032), by a grant for the professional development of new academic staff from the Ratchadapisek Somphot Fund at Chulalongkorn University, by the Thailand Research Fund (TRF), and by the Office of the Higher Education Commission (OHEC), Faculty of Science, Chulalongkorn University (RSA5980038). PB was additionally supported by a scholarship from the Royal Government of Thailand. TN was also additionally supported by a scholarship from the Development and Promotion of Science and Technology talent project (DPST).
AS was supported by a Victoria University of Wellington PhD Scholarship.
MV was supported by the Marsden Fund, via a grant administered by the Royal Society of New Zealand.
}


\end{document}